\documentstyle[preprint,prd,aps,floats,tighten,epsf]{revtex}
\begin{document}
\def\bull{\ {\vrule height 1.4ex width 1.4ex depth 0.2ex} }
\preprint{\vbox{\hbox{RUHN-00-01}}}

\title{ An alternative to domain wall fermions}

\author{Rajamani Narayanan}
\address{American Physical Society,
One Research Road, Ridge, NY 11961\\ {\tt rajamani@bnl.gov}}
\author{Herbert Neuberger}
\address{Department of Physics and Astronomy,
Rutgers University, Piscataway, NJ 08855-0849
\\ {\tt neuberg@physics.rutgers.edu}}

\maketitle
\begin{abstract}
We define a sparse hermitian lattice Dirac matrix, $H$, coupling $2n+1$
Dirac fermions. When $2n$ fermions are integrated out the induced
action for the last fermion is a rational approximation to the hermitian 
overlap Dirac operator. We provide rigorous bounds on the condition
number of $H$ and compare them to bounds for the higher dimensional
Dirac operator of domain wall fermions. Our main conclusion is that 
overlap fermions should be taken seriously as a practical alternative
to domain wall fermions in the context of numerical QCD. 
\end{abstract}

\section{Introduction}

A major embarrassment of lattice field theory in the context of
QCD has disappeared: we now have a way to preserve chiral symmetry
on the lattice ~\cite{kaplan,slav,plbfirst}, 
and the all important theoretical understanding
of ``soft'' physics consequences of chiral symmetry at the Lagrangian
level can be taken over from the continuum to the lattice. Naturally,
one is eager to exploit this development in numerical QCD, and at
the moment there are two ways that have been explored:
One is the so called domain wall fermion 
approach~\cite{kaplan,plbfirst,domain}, 
and the other is
based on implementing~\cite{ratioprl} a rational approximation to the
sign function of the overlap Dirac operator~\cite{ovldiracop}. The
relation between the two is elucidated in~\cite{trunc}. One
conclusion from~\cite{trunc} could be that with computers 
immensely more powerful than the ones we have
at present it wouldn't matter which approach one uses for numerical QCD. 
In practice, the two methods are quite different, and it is important
to assess their relative strengths and weaknesses. 

To compare methods one needs to separate
the quenched case from the dynamical case: In a dynamical hybrid
Monte Carlo simulation
one needs to invert the Dirac operator often, at each step of the so
called ``trajectory''. The inversions do not generate propagators
used in computing physical observables, except when the 
trajectory is completed and the change in gauge fields accepted. 
In a quenched simulation one extracts physics
results from each set of fermion propagators. Thus, the comparison 
of overall fractional costs for inversions works
out differently in the quenched and in the dynamical case. 

It seems that the overlap approach has some advantage in the quenched
case: it has been possible to do simulations at very low quark masses
~\cite{edwliu}, lower than what was achieved with domain wall 
fermions~\cite{columb}. 
To be sure, 
a systematic and complete comparison has not been carried out, so
this is more an impression than a hard fact. The reason for the difference
is that with the overlap it has been possible to better handle the 
cases where the Hermitian Wilson Dirac operator, $H_W$, has eigenvalues
very close to zero. Numerical investigation has shown that there is
a finite density of eigenvalues very close to zero at typical lattice
couplings used in lattice simulations~\cite{wilson}.
While this is a problem for both domain 
walls~\cite{trunc} and overlap, in the overlap case one can 
exploit the simple
local structure of $H_W$ and project out the troublesome states~\cite{scri}. 
This
is expensive, but needs to be done only once per gauge field configuration
and will help in the calculation of all propagators at this gauge field.
Thus, the extra expense is amortized in the quenched case. A similar
projection method in the case of domain wall fermions might make both
methods equal. Until very recently there existed no practical 
implementation of a projection technique
in the domain wall context, but this is changing as we write~\cite{roburs}. 

At the dynamical level it always was felt that domain wall fermions
were superior because the action was quite standard in form and one
needed only one conjugate gradient inversion, rather than the two
nested ones required by the rational overlap~\cite{ratioprl,scri}. 
Here again, this is 
more an impression than a hard fact because one ought to take into
account condition numbers, matrix sizes and the possibility to use
projectors. But, superficially at least, it is hard to ignore the
advantage of domain wall fermions with a straightforward action
and a relatively well tested algorithm. 

The latter distinction between the implementation of domain wall fermions
and rational overlap fermions could be eliminated however~\cite{fived} by
undoing what the overlap does relative to domain wall fermions. One
reintroduces extra fields which interact quadratically by a sparse 
hermitian matrix $H$.
The main requirement of $H$ is the following: There exists one 
massless field, $\psi$, such that 
integrating out all the other fields
produces for $\psi$ a specific effective action $S_{\rm eff}$.
$S_{\rm eff}=-\bar\psi H^n_o \psi$ and the approximate 
hermitian overlap Dirac operator is given by $H^n_o = {1\over 2}
(\gamma_5 + \epsilon_n (H_W ))$ where the function  $\epsilon_n (x)$
is a numerically accurate rational approximation 
to the sign function ${\rm sign} (x)$
for $x$ in the spectrum of $H_W$. One can easily add an 
explicit Dirac mass term for the field $\psi$. 

$H$ has to fulfill additional requirements: It has to have a condition number
that is not significantly worse than that of the five dimensional
Dirac operator used for domain wall fermions, $D$. In particular, 
the condition number should not diverge as a function of the number
of extra fields, $2n$. 
The cost of acting with $H$ should grow no more than linearly with $n$,
so long as the truncation of the sign function converges as $e^{-c n}$. 
Furthermore, $H$ should depend structurally
only on $H_W$ as one entity. 
A dependence on $H_W$
as one object ensures that if a better version of $H_W$ is eventually 
found, it would be trivial
to change $H$, replacing the old $H_W$ with the new one. 
Also, the implementation of the projection technique
simply involves a replacement of $H_W$ as a whole. 
In particular, taking derivatives with
respect to the gauge fields would be simple, making 
the computation of the ``force'' in hybrid Monte
Carlo relatively easy. 
Although the action for
domain wall fermions also has a simple
dependence on the gauge fields,
the dependence on the length of the extra direction is controlled by
the complicated transfer matrix $T_W$~\cite{trunc}. The dependence
on $n$ is simpler in the overlap case, as it comes in only 
through $\epsilon_n (H_W )$, and $H_W$ is a sparse matrix, unlike
$T_W$. 

Previous proposals for $H$
had some numerical entries that were large and grew with $n$. This
implied that the norm of $H$, $||H||$, grew and made the condition number
$\kappa (H) =||H||||H^{-1}||$ grow too. Our choice for matrix norms is
quite standard: $||X|| = \sqrt{\lambda_{\rm max}
{(X^\dagger X )}}$, with 
$\lambda_{\rm max} (X^\dagger X)$ being the largest 
eigenvalue of $X^\dagger X$.
The main objective of this paper is to present a new version of $H$
which satisfies the above requirements and also has a reasonable
condition number. 

In the next section we shall construct $H$. We proceed 
with the derivation of  rigorous upper bounds
on $||H||$ and on $||H^{-1}||$. Rigorous
bounds are nice, but there always is a question whether they
are saturated. This will be discussed. We shall also ask the 
opposite question: what is the best one can hope for, regarding
$\kappa (H)$ ? To answer that we shall derive a lower bound for
$||H^{-1}||$. This is all we need because, in practice, we expect
the upper bound on $||H||$ to be typically almost saturated. 
We proceed to address
the question how our results compare to what is known about the 
domain wall fermion operator $D$.
Since not much seems to be known, we derive some exact results for
$D$ too, but also leave more work for the future. We then
briefly discuss projection techniques for both overlap
and domain wall fermions. 

We conclude the paper with a discussion of other advantages 
we envisage our proposal to have over domain wall fermions, beyond
better chirality properties. 
We compare the number of $H_W$ operations
needed to perform the inversion of the fermionic operator. We find
this number numerically for three cases: domain wall fermions,
a method using the operator $H$ proposed in this paper, and a method using 
a direct computation of the sign function represented by 
the same rational approximation as implemented by $H$.
The comparison is carried out for the two dimensional Schwinger model. 
We hope to convince the reader that the overlap alternatives merit
serious numerical testing in the context of QCD, something we are not fully
geared up to do efficiently by ourselves.

\section{Construction of $H$}

The kernels of the quadratic fermionic actions we shall work with will
be hermitian matrices in most cases. Let $\psi$ be the light Dirac
field representing a light quark. We wish to end up with an action
\begin{equation}
S_{\rm eff} (\psi) = - \bar\psi \left [
{{1+\mu}\over 2} \gamma_5 + 
{{1-\mu}\over 2}\varepsilon_n (H_W ) \right ]\psi .
\end{equation}
The bare quark mass $\mu$~\cite{trunc} is restricted by $|\mu| <1$ for
physical reasons. 

The matrix $H_W$ has the standard Wilson form, 
but can be easily replaced by a more elaborate construction.
The function $\varepsilon_n (x)$ is an approximation to ${\rm sign} (x)$:
\begin{equation}
\varepsilon_n (H_W ) = {1\over n } \sum_{s=1}^n {1\over {c_s^2 H_W +
{s_s^2 \over H_W}}}
\end{equation}
with
\begin{equation}
c_s= \cos \theta_s ,~~s_s=\sin \theta_s,~~~\theta_s =
{\pi\over{2n}} (s-{1\over 2}),~s=1,2,...,n .
\end{equation}
This rational approximation can be replaced by others:
One can replace the $c_s^2$ 
and $s_s$ quantities with other real
numbers of either sign; 
to change the overall sign of a contribution at some $s$ one 
simply switches the sign of $H_W$ at that $s$. 

For each $s$ we introduce a new field $\chi_s$. The $\chi_s$
fields are decoupled from each other and enter quadratically
in the action: 
\begin{equation}
\bar \chi_s \left [ c_s^2 H_W + {s_s^2 \over H_W } \right ] \chi_s .
\end{equation}
In addition, they couple to the $\psi$ field by
\begin{equation}
\sqrt{{1-\mu}\over {2n}} \sum_s (\bar\psi \chi_s + \bar \chi_s \psi ) .
\end{equation}
To get the right effective action for $\psi$, there also is a 
quadratic term in $\psi$:
\begin{equation}
-{{1+\mu}\over 2}\bar\psi \gamma_5 \psi .
\end{equation}

To make the action local we should eliminate the inverse of
$H_W$ from the $\chi$-action. To this end we introduce $n$ more
Dirac fields, $\phi_s$, also Grassmann, and change the $\chi$ self
interaction to:
$$
\bar \chi_s ( c_s^2 H_W ) \chi_s +s_s ( \bar \chi_s \phi_s +
\bar \phi_s \chi_s ) - \bar\phi_s (H_W ) \phi_s $$

Introduce the combined field 
$\bar\Psi = (\bar\psi, \bar\chi_1 ,\bar\phi_1,...,\bar\chi_n,\bar\phi_n )$. 
The action is
\begin{equation}
S= \bar \Psi H \Psi
\end{equation}
with
\begin{equation}
H=\pmatrix{ 
-{{1+\mu}\over 2} \gamma_5 & \sqrt{{1-\mu}\over {2n}}&0&\sqrt{{1-\mu}\over
{2n}}&0&\dots
&\sqrt{{1-\mu}\over {2n}}&0\cr
\sqrt{{1-\mu}\over {2n}}& c_1^2 H_W & s_1 &0 &0 &\dots &0&0\cr
0& s_1 &-H_W &0&0 &\dots &0&0\cr
\sqrt{{1-\mu}\over {2n}}&0&0&c_2^2 H_W & s_2 &\dots& 0& 0\cr
0&0&0&s_2 &-H_W &\dots&0&0\cr
\vdots& \vdots& \vdots& \vdots& &\vdots&0&0\cr
\sqrt{{1-\mu}\over {2n}} &0&0&0&0&\dots&c_n^2 H_W & s_n\cr
0&0&0&0&0&\dots&s_n&-H_W\cr
}
\end{equation}

Our new extended overlap model is based on the following identity:
\begin{equation}
\int d\bar\Psi d\Psi e^{\bar\Psi H \Psi} = 
\left [ \prod_{s=1}^n \det (c_s^2 H_W^2 +s_s^2 )
\right ] \int d\bar\psi d\psi e^{-\bar\psi \left ( {{1+\mu}\over 2}
\gamma_5 +{{1-\mu}\over {2n}}\sum_{s=1}^n
{1\over {c_s^2 H_W + {{s_s^2}\over H_W}}} \right ) \psi }
\end{equation}
The prefactor can be canceled by adding pseudofermions, which will
be decoupled in the $s$ index. 

At $n=\infty$ we can take a continuum limit, with
\begin{equation}
d\theta = {\pi\over{2n}} ds ,\ \ \ 0<\theta <{\pi\over 2}, \ \ \
\chi_s ={1\over\sqrt{n}} \chi(\theta ), \ \ \ \phi_s = {1\over\sqrt{n}} 
\phi(\theta ) .
\end{equation}
The effective action then becomes:
\begin{eqnarray}
S= &-&{{1+\mu}\over 2} \bar\psi\gamma_5 \psi +
{2\over\pi} \sqrt{{{1-\mu}\over 2}}
[\bar\psi \int d\theta \chi(\theta ) + \int d\theta \bar\chi (\theta ) \psi ]
\cr
&+&{2\over\pi} \int d\theta [\cos^2 \theta  \bar\chi (\theta ) H_W \chi (\theta)
-\bar\phi (\theta )H_W \phi (\theta ) +
\sin\theta \bar \chi (\theta ) \phi (\theta ) + 
\sin\theta \bar\phi(\theta )\chi(\theta) ] .
\end{eqnarray}
The following identity (which holds as long as $||H_W^{-1}||$ is finite)
expresses the essence of our construction: 
\begin{equation}
{2\over\pi} \int_0^{\pi\over 2} 
{{d\theta}\over{H_W \cos^2 \theta  + {{\sin^2 \theta  }\over {H_W}}}}=
{2\over\pi} \int_0^\infty {{dt}\over t}
{1\over{{{H_W}\over t} + {t\over{H_W}}}}
={\rm sign} (H_W ) .
\end{equation}
In terms of $\theta$ the action in the extra dimension
involves no derivatives. In a rough sense the $\theta$ variable
corresponds to a ``fifth'' momentum (for a four dimensional physics
application). 
The physical degrees of freedom, $\psi$, couple
only to an average field. These two features indicate that one
might be able to use some sort of multi-grid or hierarchical techniques to
reduce the ${\cal O} (n)$  computational cost per inversion
of $H$ to a logarithmic dependence on $n$. 
Other integral representations of the sign function, or other approximation
to the integrals above, will produce more variations on the same
basic idea, but with possibly different computational properties.

The strategy for finding bounds on the spectrum of $H$ is based on
an exact formula for the determinant of $H-z$. This determinant is
obtained by adding $-z\bar\Psi \Psi$ to the action and doing
the gaussian integral.
\begin{equation}
\det (H-z) = \left \{ \prod_{s=1}^n 
\det \left [ (c_s^2 H_W -z )(H_W +z ) +s_s^2 \right ] \right \}
\det \left [ {{1+\mu}\over 2}\gamma_5 +z 
+{{1-\mu}\over 2} f_n (H_W ,z )\right ]
\end{equation}
Here,
\begin{equation}
f_n (H_W , z ) = {1\over n} \sum_{s=1}^n {1\over {c_s^2 H_W +
{s_s^2 \over {H_W +z }} -z}} .
\end{equation}
Eigenvalues of $H$ are roots of the equation $\det (H-z)=0$.
All the roots come from roots of the last factor. (Roots of the
factors in the product over $s$ are canceled by poles in the last
factor. So, the spectrum of $H$ is determined by the last factor.) 

Introducing
\begin{equation}
S_n(a,b) = {1\over n} \sum_{s=1}^n {1\over {a c_s^2 +b s_s^2}},
\end{equation}
we see that
\begin{equation}
f_n (H_W , z ) = S_n(H_W -z , {1\over{H_W +z }} -z ).
\end{equation}

$S_n(a,b)$ is a ratio of polynomials in $a,b$:
\begin{equation}
S_n(a,b) = {{P_n(a,b)}\over {Q_n(a,b)}},~~~P_n(a,b)=\sum_{s=1}^n 
{{2n}\choose{2s-1}} b^{n-s} a^{s-1},~~~Q_n(a,b)=\sum_{s=0}^n
{{2n}\choose{2s}} b^{n-s} a^s .
\end{equation}
Extracting $(H_W +z)^{-n}$ from numerator and denominator, we see that
$S(a,b)$ is also a ratio of polynomials in $H_W$. 
For real $a$ and $b$ we can write simple closed formulae:
If $ab >0$ we have
\begin{equation}
S_n(a,b) = {{{\rm sign} (b)}\over \sqrt{ab}} \tanh (n\omega),~~~
\omega = \log\left ( |1+\sqrt {a/b} | /|1-\sqrt {a/b }| \right ) .
\end{equation}
If $ab <0$,
\begin{equation}
S_n(a,b) = {{{\rm sign} (b)}\over \sqrt{-ab}} \tan (n\omega),~~~
0\le \omega \le {\pi\over 2},~~ 
e^{i\omega} = ( 1+i\sqrt {-a/b} )/(1-i\sqrt{-a/b}) .
\end{equation}
These formulae make the $n$-dependence explicit. 

The product $ab$ clearly plays a central role:
\begin{equation}
ab=1+zg(H_W ,z),~~~~g(H_W,z) = z-H_W -{2\over{H_W +z}} .
\end{equation}

Our bounds are based on the observation that for $ab >0$
we have an $n$-independent bound:
\begin{equation}
|f_n (H_W, z)|=|S_n(a,b)|\le {1\over \sqrt {ab}} .
\end{equation}
If we allowed $ab <0$, we would have had 
little control over $S_n(a,b)$ because of the
tangent function.

\section{ Upper bound on $||H||$}

For very large $|z|$ $ab$ is large and positive, making
$z+{{1-\mu}\over 2}
f_n (H_W ,z)$ large in absolute value. It is clear 
that $\det (H-z)$ cannot vanish then. To find
an upper bound for $||H||$ we look for the smallest $|z|$ values
for which the above is still true, as we decrease $|z|$ from infinity.

\noindent {\bf Theorem I:} For any hermitian bounded $H_W$ we have
\begin{equation}
||H||\le \sqrt {||H_W||^2 +2 } .
\end{equation}
\noindent {\bf Proof of Theorem I:} 
Pick a real number $z$,
\begin{equation}
|z| \ge \sqrt{||H_W||^2 +2 } .
\end{equation}
We shall prove, by contradiction, that this implies $\det (H-z)\ne 0$.
All we need to show is that the hermitian operator 
${{1+\mu}\over 2} \gamma_5 + z+{{1-\mu}\over 2} f_n (H_W , z)$
has no zero eigenvectors. 
We assume that there exists such a normalized eigenvector 
$\psi_0$. $\psi_0$ obeys:
\begin{equation}
{{1+\mu}\over 2} \gamma_5\psi_0= -\left [
z+ {{1-\mu}\over 2} f_n (H_W , z)\right ] \psi_0 .
\end{equation}
Taking norms we get
\begin{equation}
{{1+\mu}\over 2} = \sqrt {\psi_0^\dagger \left [
z+ {{1-\mu}\over 2} f_n (H_W , z)\right ]^2 \psi_0 } .
\label{upper_bound}
\end{equation}

Let $h$ be an arbitrary eigenvalue of $H_W$; we know that
\begin{equation}
|z|\ge \sqrt{h^2 +2} .
\end{equation}
Simple analysis of the function $zg(h,z)$ implies
\begin{equation}
zg(h,z) \ge 0 .
\end{equation}
As a consequence:
\begin{equation}
|f_n (h,z )| \le 1 .
\end{equation}
Hence, every eigenvalue $\lambda$ of $z+{{1-\mu}\over 2} f_n (H_W , z)$
obeys
\begin{equation}
|\lambda| \ge |z|-{{1-\mu}\over 2} .
\end{equation}
(Recall that $|\mu|\le 1$.) By the variational principle, the right hand
side of equation~(\ref{upper_bound}) obeys the same inequality. Hence,
\begin{equation}
{{1+\mu}\over 2} \ge |z| - {{1-\mu}\over 2}
\end{equation}
which implies $|z|\le 1$ in contradiction to our initial assumption
about $z$ $\bull$ 

This establishes the upper bound of Theorem I. 
Note that it is $\mu$ and $n$ independent.

\noindent {\bf Corollary:} Let $H_W$ be the hermitian Wilson Dirac
operator in even $d$ dimensions, and $H$ defined as above, with
the standard replacement of the four dimensional $\gamma_5$. Let
the mass parameter $m$ in $H_W$ be restricted by $m>-2$. Then
\begin{equation}
||H|| \le \sqrt{2 + (2d+m)^2} .
\end{equation}
This is a direct consequence of the known upper bound on $||H_W||$ $\bull$

We expect this bound to be quite close to optimal. 

This proves that the main difficulty faced by previous proposals
to implement the overlap Dirac operator by adding extra fields has
been completely eliminated. The problem with the previous proposals
was that one could not control $||H||$. Moreover, our bound is
quite stringent numerically, even smaller than a typical bound on
a Wilson Dirac operator in $2d+1$ dimensions ~\cite{bounds}. 
This is a plausible
comparison, because one could thing about $s$ as
indexing an extra dimension. At any rate, for $d=4$ and $m=-1.8$
one would have 
\begin{equation}
||H||\le 6.4
\end{equation}
which is very reasonable. 

We ran some tests in two dimensions with $m=-1$ and found the bound
of $3.3$ to be typically almost saturated for $U(1)$ 
gauge configurations generated
with a Wilson action at $\beta =2$. Moreover, with a trivial 
gauge background one can explicitly check that one can get
quite close numerically to the upper bound and 
so it is impossible to find a gauge background
independent bound which is significantly better. 

\section{Upper bound on $||H^{-1}||$}

The  immediate question we need to address now is what happens
to the low eigenvalues of $H$. Having established that $H$ behaves
more or less as a usual fermionic lattice operator at high eigenvalues,
numerical problems can only come from a large $||H^{-1}||$. 

The basic strategy to get the bound is similar to the one used
above. First take $z=0$ and $n=\infty$. Assume that
$H_W$ has no zero eigenvalue. Then, $ab=1$ and 
$||f_\infty (H_W, 0)||=1$. As a result, ${{1+\mu}\over 2}\gamma_5 +
{{1-\mu}\over 2} f_\infty (H_W, 0) $ cannot have zero eigenvalues
if $\mu \ne 0$. We therefore look now for a neighborhood of $z=0$
where, if $\mu >0$ and $H_W$ has no zero mode, we have
\begin{equation}
{{1+\mu}\over 2} > ||z +{{1-\mu}\over 2} f_n (H_W, z)|| .
\end{equation}

If this is true, ${{1+\mu}\over 2} \gamma_5 +z+ 
{{1-\mu}\over 2}f_n (H_W , z )$ cannot have a zero mode.
Indeed, if there were such a zero mode, 
$\psi_0$, we would have a contradiction, since
then 
\begin{equation}
\left ( {{1+\mu}\over 2}\right )^2
 \psi_0^\dagger \psi_0 = \psi_0^\dagger \left [ z+{{1-\mu}\over 2}
f_n (H_W , z)\right ]^2 \psi_0 ,
\end{equation}
in violation of the variational principle for the maximal
eigenvalue of ${\left [ z+{{1-\mu}\over 2}f_n (H_W , z ) \right ]}^2$.

The range $0<\mu <1$ covers all possible positive quark masses.
The case $-1 < \mu < 0$ describes quarks with a negative mass.
This certainly is not without interest, but the analysis
becomes more complicated and unnecessary for our 
purposes here. So, we simply restrict ourselves to the range $0<\mu<1$.

Let $h$ be an eigenvalue of $H_W$:
\begin{equation}
{1\over{||H_W^{-1}||}} \le |h|\le ||H_W|| .
\end{equation}
We define $z_n (h, \mu )$ as the smallest positive solution to
the following equation:
\begin{equation}
{{1+\mu}\over 2} = | z_n (h,\mu) +{{1-\mu}\over 2} f_n (h, z_n (h,\mu))| .
\label{small}
\end{equation}
Since $f_n (-h, -z ) = -f_n (h,z)$, we have:
\begin{equation}
z_n (h,\mu ) = - z_n (-h , \mu) .
\end{equation}
$h$ is restricted to ranges that are symmetric about zero. Therefore,
the inequality
\begin{equation}
{{1+\mu}\over 2} > | z +{{1-\mu}\over 2} f_n (h, z)|
\end{equation}
can be guaranteed by restricting $z$ to a range symmetric about zero:
\begin{equation}
|z| < u_n (H_W , \mu ) .
\end{equation}
Here, the numerical bound $u_n (H_W ,\mu )$ is given by
\begin{equation}
u_n (H_W , \mu ) = \min_{{1\over {||H_W^{-1}||}} 
\le |h| \le ||H_W||} z_n (h, \mu) .
\label{unhw}
\end{equation}
We have therefore
proven the following bound: 

\noindent {\bf Theorem II:} For $0 \le \mu < 1$ and arbitrary nonsingular
hermitian $H_W$, we have:
\begin{equation}
{1\over{||H^{-1}||}} \ge u_n (H_W , \mu ) .
\end{equation}
$u_n (H_W , \mu )$ is determined by equations~(\ref{small}) 
and~(\ref{unhw}) $\bull$ 

$z_n (h, \mu )$ can be found numerically for the ranges 
of $h$ and $\mu$ of interest in practice.
To find a bound analytically is cumbersome because $u_n (H_W, \mu)$
will be controlled by either an $h$ corresponding to a maximal
(in absolute magnitude) eigenvalue of $H_W$ or to a minimal one. Some
analysis shows that there is a constant of order unity
(dependent on the
dimension $d$ and the $m$ parameter in $H_W$), $C$, 
and that if we restrict $H_W$ by
\begin{equation}
||H_W^{-1}|| \ge C ,
\end{equation}
the minimum over $h$ we need in
order to get $u_n (H_W , \mu )$ is attained at 
\begin{equation}
h=h_{\min}={1\over {||H^{-1}||}} .
\end{equation}
Thus, the entire dependence of $u_n (H_W , \mu )$ 
on $H_W$ comes in through the value
of the lowest eigenvalue of $H_W^2$.
Gauge configurations for which $||H_W^{-1}|| < C$ are easily handled
by any numerical method and the condition number in that case is
not a source of concern. So, nothing is lost by assuming that
the smallest eigenvalue of $H_W^2$ is smaller than $C^{-2}$; this
is anyhow the case for most gauge configurations one encounters
in practical QCD simulations.

In order to get some feeling for orders of magnitude we work
out a simple analytical approximation for the bound on $|z|$ and
represent the exact result for the bound as the product of this
approximation times a correction factor. Some typical values for the
correction factor are presented in Table~\ref{tab_unhw}. 
All the entries in Table~\ref{tab_unhw}
are for the case when the smallest eigenvalue of $H_W^2$ 
is smaller than $C^{-2}$.

We now take $ab >0$ because this certainly is true at $z=0$.
As explained above, we can restrict our attention to positive $z$,
without loosing generality. It is true now that
\begin{equation}
||z+{{1-\mu}\over 2} f_n (h_{\rm min} , z)||\le z+
{{1-\mu}\over 2} {1\over \sqrt{1+zg(h_{\rm min} , z )}} .
\end{equation}
So, our bound will hold if we enforce the first inequality below:
\begin{eqnarray}
{{1+\mu}\over 2} 
&>& z +{{1-\mu}\over 2} {1\over\sqrt{1+zg(h_{\rm min} , z )}}
\approx z+{{1-\mu}\over 2} \left [ 1+ z\left ( {h_{\rm min} \over 2} + 
{1\over h_{\rm min}} \right ) \right ] \cr
&\approx& z+ {{1-\mu}\over 2} +
{{1-\mu}\over 2} {z\over h_{\rm min} }\approx {{1-\mu}\over 2} 
+{z\over{2h_{\rm min}}} .
\end{eqnarray}
In the above series of approximations we assumed $h_{\rm min} <<1$.
We end up with an approximate range from which eigenvalues of $H$
are excluded:
\begin{equation}
|z| < 2\mu h_{\rm min} .
\end{equation}

Finally, we rewrite our exact result in the following form:
\begin{equation}
{1\over{||H^{-1}||}} \ge u_n (H_W , \mu ) \equiv c_n (H_W , \mu ) 
{{2\mu}\over{||H_W^{-1}||}} .
\end{equation}
For $||H_W^{-1}|| \ge C$ the $c_n$ prefactor depends on $H_W$
just through $h_{\rm min}$. In Table~\ref{tab_unhw} we collected several
values of $c_n (h_{\rm min} , \mu )$ to give a sense of the
dependence on $n$, $h_{\rm min}$ and $\mu$. $c_n (H_W , \mu )$
is calculated using equations (36), (40) and (47).

\begin{table}
\begin{tabular}{ccccccc}
    $h$  & $\mu$ &   $n=10$   &   $n=20$   &   $n=50$   &   $n=100$  &
$n=\infty$ \cr
\hline
    .01  &  .01  &  158.9903  &   68.2685  &   15.1871  &    2.7847  &
.9804  \cr
    .01  &  .02  &   80.9208  &   34.9814  &    8.1349  &    1.8857  &
.9803  \cr
    .01  &  .04  &   41.8905  &   18.3404  &    4.6106  &    1.4366  &
.9796  \cr
    .01  &  .08  &   22.3825  &   10.0234  &    2.8506  &    1.2111  &
.9759  \cr
    .01  &  .16  &   12.6341  &    5.8657  &    1.9686  &    1.0892  &
.9600  \cr
    .02  &  .01  &   65.9303  &   22.8212  &    2.7302  &     .9939  &
.9616  \cr
    .02  &  .02  &   33.7907  &   11.9821  &    1.8492  &     .9779  &
.9617  \cr
    .02  &  .04  &   17.7244  &    6.5649  &    1.4094  &     .9696  &
.9614  \cr
    .02  &  .08  &    9.6970  &    3.8598  &    1.1892  &     .9630  &
.9587  \cr
    .02  &  .16  &    5.6891  &    2.5087  &    1.0718  &     .9478  &
.9451  \cr
    .04  &  .01  &   21.7717  &    4.7348  &     .9567  &     .9258  &
.9258  \cr
    .04  &  .02  &   11.4341  &    2.8394  &     .9417  &     .9262  &
.9262  \cr
    .04  &  .04  &    6.2683  &    1.8934  &     .9344  &     .9266  &
.9266  \cr
    .04  &  .08  &    3.6903  &    1.4224  &     .9297  &     .9255  &
.9255  \cr
    .04  &  .16  &    2.4066  &    1.1852  &     .9187  &     .9162  &
.9162  \cr
    .08  &  .01  &    4.3730  &    1.0017  &     .8606  &     .8606  &
.8606  \cr
    .08  &  .02  &    2.6252  &     .9322  &     .8614  &     .8614  &
.8614  \cr
    .08  &  .04  &    1.7535  &     .8985  &     .8628  &     .8628  &
.8628  \cr
    .08  &  .08  &    1.3211  &     .8826  &     .8641  &     .8641  &
.8641  \cr
    .08  &  .16  &    1.1076  &     .8719  &     .8612  &     .8612  &
.8612  \cr
    .16  &  .01  &     .8695  &     .7517  &     .7515  &     .7515  &
.7515  \cr
    .16  &  .02  &     .8118  &     .7528  &     .7527  &     .7527  &
.7527  \cr
    .16  &  .04  &     .7846  &     .7549  &     .7549  &     .7549  &
.7549  \cr
    .16  &  .08  &     .7738  &     .7586  &     .7586  &     .7586  &
.7586  \cr
    .16  &  .16  &     .7714  &     .7629  &     .7629  &     .7629  &
.7629  \cr
\end{tabular}
\vspace{5mm}
\caption
{\sl The entries are numerically obtained
exact values for the coefficient $c_n (h, \mu )$
appearing
in our rigorous bounds. $h^2$ always is the smallest eigenvalue of $H_W^2$.}
\label{tab_unhw}
\end{table}

\section{Condition number of $H$: worst and best case}
 
Combining our exact results 
we obtain an exact bound for condition numbers.

\noindent{\bf Theorem III}: 
\begin{equation}
\kappa (H) \le {{\kappa (H_W )} \over {2\mu}} 
{{\sqrt {1 +{2\over {||H_W||^2}}}} \over 
{c_n (H_W, \mu )}
}
\label{cond_nr_ineq}
\end{equation}
\rightline{$\bull$}

{}From Table~\ref{tab_unhw}
we see that decreasing $n$ improves the condition number.
This is expected because finite $n$ effects induce chirality
violation even at $\mu=0$, so act somewhat as an effective additional mass. 
We also see from 
Table~\ref{tab_unhw} that at high enough $n$ the entire correction
needed to make the approximate bound rigorous is of order unity.
The dependence of the correction factor on $h$ and $\mu$ 
shows that the approximation is somewhat pessimistic. 

In summary, 
roughly, the main message is that the condition number of 
$H$ cannot be worse than  the product
of the condition numbers for $H_W$ and for the effective action
governing the light fermions, 
\begin{equation}
\kappa \left ( {{1+\mu}\over 2} \gamma_5 + {{1-\mu}\over 2} \varepsilon
(H_W ) \right ) \sim {1\over \mu} .
\end{equation}

It is natural to ask now: How pessimistic is the bound when viewed as an 
estimate for $\kappa (H)$ ? We shall answer this question in two ways:
First, we shall find an upper bound to ${1\over{||H^{-1}||}}$. Since
the upper bound we found for $||H||$ is a good estimate, this would
provide us with a best possible condition number. Next, we shall show 
that one cannot rule out a background gauge field configuration for
which the lower bound on ${1\over{||H^{-1}||}}$ is saturated. These
two results put $\kappa(H)$ into a range. The remaining practical 
question is where in the range will $\kappa(H)$ typically be. The
answer to this question will depend on details of the pure gauge action,
and on whether we are dealing with a quenched simulation or with a dynamical
one. 

Upper bounds for the smallest eigenvalue of $H^2$ are easily obtained
from the variational principle. Look at all the block diagonal elements
of $H^2$:
\begin{equation}
H^2_{\psi\psi} = {{3+\mu^2}\over 4},~~H^2_{\chi_s \chi_s}= c_s^4 H_W^2 +s_s^2 +
{{1-\mu}\over {2n}},~~H^2_{\phi_s \phi_s} = H_W^2 + s_s^2 .
\end{equation}
Imagine computing the expectation value of $H^2$ in a state with an
eigenstate of $H_W$ with eigenvalue $h$ 
in one block, and zero for all other blocks. The variational
principle implies then Theorem IV.

\noindent{\bf Theorem IV:}
\begin{equation}
\lambda_{\rm min} (H^2) \le \min_h \left \{ {{3+\mu^2}\over 4}, 
\min_{s} \left ( c_s^4 h^2 +
s_s^2 + {{1-\mu}\over {2n}} \right ), \min_s \left ( h^2 + s_s^2 \right ) \right
\}
\end{equation}
\rightline{$\bull$}

We assume that $h_{\rm min} \le {1\over{\sqrt{2}}}$. Then, the minimum
is not attained in the $\psi\psi$ block. In the other blocks,
the best is to take $h=h_{\rm min}$. Among the $\chi_s \chi_s $ blocks 
and 
the $\phi_s \phi_s $ blocks the minimum occurs at $s=1$. Thus we find:
\begin{equation}
\lambda_{\rm min} (H^2) \le {\rm min} \left \{ \left ( (1-\sin^2 {\pi\over {4n}}
)^2
h_{\rm min}^2 + \sin^2 {\pi\over {2n}} + {{1-\mu}\over {2n}}\right ),
\left ( h_{\rm min}^2 +\sin^2 {\pi\over {2n}}\right )\right \} .
\end{equation}
In the physical relevant cases, $\mu$ will be small enough and $n$ will
be large enough to give:
\begin{equation}
{1\over{||H^{-1}||}} \le \sqrt{ 
{1\over{||H_W^{-1}||^2}} +\sin^2 {\pi\over{2n}} }.
\label{iva}
\end{equation}
To get a good approximation for the sign function we need $n
>>{\rm max} \left \{ ||H_W||, {1\over{||H_W^{-1}||}} \right \}
$, so 
\begin{equation}
\sqrt{ 
{1\over{||H_W^{-1}||^2}} +\sin^2 {\pi\over{2n}} }\approx 
{1\over{||H_W^{-1}||}}
\label{ivb}
\end{equation}

{}From this we learn that, roughly, the condition number of $H$ cannot
be better than the condition number of $H_W$. If we use a low value of $n$,
for which the approximation to the sign function is bad, and hence when
there are significant violations of chirality 
beyond the explicit mass term $\mu$, the condition number of $H$ might
be better than that of $H_W$. This situation however probably defeats
the purpose of using the new fermions instead of the older, more economical
approach of fine tuning the mass term in $H_W$ to small effective quark mass. 

On physical grounds it is obvious that there also must exist an upper
bound on the lowest eigenvalue of $H^2$ in terms of the ``bare'' quark mass
$\mu$. To see this we recall that the basic identity of the entire
approach we are focusing on in this paper is
\begin{equation}
\left ( H^{-1}\right )_{\psi\psi} = {1\over { {{1+\mu}\over 2}\gamma_5 + 
{{1-\mu}\over 2} \varepsilon_n (H_W )}} .
\end{equation}
The variational principle then gives us Theorem V.

\noindent{\bf Theorem V:}
\begin{equation}
\lambda_{\rm min} (H^2 ) \le \lambda_{\rm min} 
\left [   {{1+\mu}\over 2}\gamma_5 + 
{{1-\mu}\over 2} \varepsilon_n (H_W ) \right ]^2 .
\end{equation}

\noindent{\bf Proof of Theorem V:} 
Let $\Phi=(\psi_0,0,0,...)$.
On very general grounds, we have
\begin{equation}
{1\over{\lambda_{\rm min} (H^2 )}} \ge
\langle \Phi | H^{-2} |\Phi \rangle 
\ge \langle \Phi | H^{-1} | \Phi \rangle^2 =
\langle \psi_0 (H^{-1})_{\psi\psi} \psi_0 \rangle .
\end{equation}
Choosing $\psi_0$ as the eigenstate of $ \left ( {{1+\mu}\over 2}\gamma_5 + 
{{1-\mu}\over 2} \varepsilon_n (H_W )\right ) $ of lowest eigenvalue 
in absolute value establishes the inequality${\bull}$

Note that this inequality is a direct
consequence of the existence of some light fermion in the theory
described by $H$. Thus, this inequality is very general. 

One cannot exclude backgrounds for which $\varepsilon_\infty (H_W )$
and  $\gamma_5$ have a common eigenvector with the eigenvalues of opposite
sign. Such eigenvectors should occur in instanton backgrounds, for example.
For such a background the right hand side of the above equation,
which is always bounded from below by $\mu$ (assume $\mu \ge 0$, as before)
practically saturates the bound. In summary, we can say that 
there are gauge field backgrounds for which we know for sure that 
$\lambda_{\rm min} (H^2) \le \mu^2$. Note that while any eigenvector
of $H_W$ is also an eigenvector of $\varepsilon_\infty (H_W )$ the
opposite is not true. 

We see that one cannot hope that the condition number
of $H$ would be smaller than either the condition number of
$H_W$, or than a number of order ${1\over\mu}$:
$\kappa(H) \ge {\rm max} \{ \kappa(H_W ), {1\over\mu}\}$. The upper bound 
on $\kappa (H)$, which we
derived previously (equation~(\ref{cond_nr_ineq})), said that
the condition number of $H$ cannot be worse (larger) 
than, roughly, ${{\kappa(H_W)}\over \mu}$. 

The last question is whether we can imagine a gauge background for which
the pessimistic (upper) bound (equation~(\ref{cond_nr_ineq})) on 
$\kappa (H)$ is saturated. 
The answer is that one
can. If the eigenstate of $H_W$ with eigenvalue closest to zero
also is an eigenstate of $\gamma_5$ the worst case will be realized.
In practice however this will happen rarely. The most common
low eigenstates of $H_W^2$, which are the main source of difficulties
in simulations, are typically non-degenerate and far from chiral 
(as can be checked by computing the
expectation value of $\gamma_5$, which would be $\pm 1$ for a chiral state).
Therefore, in simulations with typical parameters employed today, one
does not expect to often realize the worst case. 

It is of course important to see how these analytic considerations
apply to practical simulations. As a toy model we simulated 2-dimensional
QED with gauge coupling $\beta=4$ on an $8\times 8$ lattice. The main
factor governing convergence was the minimal eigenvalue in absolute
value of $H$.
We found that the upper bound of equation~\ref{cond_nr_ineq}, when
viewed as an estimate, is overly pessimistic by
a factor of order 2. 

\section{Domain wall fermions}

For domain wall fermions we have, in total, $n$ Dirac fermions labeled
by $s$. The light fermion is not as sharply identified as before. Its
left handed and right handed components reside predominantly at $s=1$ and
at $s=n$. The action is $\bar\Psi D \Psi$, where, adopting notation
from~\cite{plbfirst}, we have in four dimensions
\begin{equation}
D=D_W -P_L (M-1)  - P_R (M -1)^\dagger ,
\ \ \ P_L={{1+\gamma_5 }\over 2}, ~P_R ={{1-\gamma_5 }\over 2} .
\end{equation}
$D_W$ is the Wilson Dirac operator with mass $m$ in the range $(-2,0)$, as
before. $D_W$ is unity in $s$-space. $M-1$ and $(M-1)^\dagger$ have nontrivial
action only in $s$-space, where they approximate first order derivatives
plus a mass term coupled predominantly to the lightest fermion.
\begin{equation}
M=\pmatrix { 0&1&0&0&0&\dots&0\cr
             0&0&1&0&0&\dots&0\cr
             \vdots&\vdots&\vdots&\vdots&\vdots&\dots&\vdots\cr
             0&0&0&0&0&\dots&1\cr
             -\mu &0&0&0&0&\dots& 0\cr}
\end{equation}
$D$ has a hermitian version, $H_{\rm dwf}$. 
Let the hermitian matrix $S$ produce
a flip in $s$-space ($S^2=1$):
\begin{equation}
S=\pmatrix{0&0&0&\dots&0&1\cr
           0&0&0&\dots&1&0\cr
          \vdots&\vdots&\vdots&\dots&\vdots&\vdots\cr
           0&1&0&\dots&0&0\cr
           1&0&0&\dots&0&0\cr}
\end{equation}
Then, with $H_W=\gamma_5 D_W$, we define
\begin{equation}
H_{\rm dwf}=\gamma_5 S D= S H_W -P_L M_L - P_R M_R ,
\end{equation}
leading to:
\begin{equation}
D^\dagger D = H_{\rm dwf}^2 .
\end{equation}
The hermitian mass operators $M_R = S(M-1)$ and $M_L = S(1-M^\dagger)$
are related by $SM_R = -M_L S$.

Following~\cite{trunc} we can derive a 
closed formula for $\det (D^\dagger D -z^2)$,
but the analysis gets complicated and we are 
not sure that it can be completed. 
Therefore, we leave this issue for the future,
and restrict ourselves here to just deriving an upper bound to
$\lambda_{\rm min} (D^\dagger D)$. Since $\lambda_{\rm max} (D^\dagger D)$
would behave similarly to $D_W^\dagger D_W$ 
in one dimension higher, we know that
$||D||$ will be of the order $10$ for QCD~\cite{bounds}. Thus, an upper bound
on ${1\over{||D^{-1}||}}$, will effectively provide a best case
for $\kappa (D^\dagger D)$. Our aim is to show that this best case
is similar to what we have found for the model we analyzed in the previous
sections.

Let $\psi_0$ be a normalized eigenstate
of $H_W$ which has the smallest eigenvalue as an eigenstate of $H_W^2$.
Construct a trial state in the larger space in which $D$ operates:
\begin{equation}
\Psi = {1 \over \sqrt{n}}\pmatrix {\psi_0\cr\psi_0\cr\vdots\cr\psi_0\cr}
=S\Psi .
\end{equation}
We find:
\begin{equation}
H_{\rm dwf} \Psi = \pm \lambda_{\rm min}^{1/2} (H_W^2 ) \Psi
+{{1+\mu}\over\sqrt{n}} 
\pmatrix {-P_R \psi_0\cr 0 \cr\vdots\cr 0\cr
P_L \psi_0\cr}
\end{equation}
We now take the norm
\begin{equation}
||D\Psi||^2 = \lambda_{\rm min} (H_W^2 )  + {{(1+\mu)^2}\over n} 
\pm 2 {{1+\mu}\over n} \lambda_{\rm min}^{1/2} (H_W^2) 
\psi_0^\dagger \gamma_5 \psi_0 .
\end{equation}
This leads to:
\begin{equation}
||D\Psi||^2 \le \lambda_{\rm min}(H_W^2 ) + {{(1+\mu)^2}\over n}
+2{{1+\mu}\over n} \lambda_{\rm min}^{1/2} (H_W^2 ) .
\end{equation}
Our final bound is given by Theorem VI:

\noindent{\bf Theorem VI:}
\begin{equation}
\lambda_{\rm min} (D^\dagger D) \le \left [ \lambda_{\rm min}^{1/2} (H_W^2 )
+{{1+\mu}\over n} \right ]^2 +(1+\mu)^2 \left ({1\over n} -{1\over {n^2}}
\right )
\label{vi}
\end{equation}
\rightline{$\bull$}

For domain wall fermions to really represent massless quarks
when $\mu=0$ we need~\cite{trunc}
\begin{equation}
n\lambda_{\rm min}^{1/2} (H_W^2 ) >>1 .
\end{equation}
If we also make the stronger assumption (stronger because we loose
nothing by treating only the case $\lambda_{\rm min} (H_W^2 ) <1$)
\begin{equation}
n\lambda_{\rm min} (H_W^2 ) >>1,
\end{equation}
the right hand side of the bound in equation~(\ref{vi}) becomes
just $\lambda_{\rm min} (H_W^2 )$ which is the same as in the previous
analysis, see equations~(\ref{iva},\ref{ivb}). 
The difference is in the finite $n$ effects: they
are larger here. This could be something that works in favor
of domain wall fermions. Of course, there is no guarantee that the
best case, analyzed here, is a good approximation to the typical
case. One can also analyze trial states with some structure in 
$s$-space:
\begin{equation}
\Psi =\pmatrix {
{c_1\psi_0} \cr {c_2\psi_0}\cr 
{c_3\psi_0} \cr \vdots \cr {c_n \psi_0} \cr }
\end{equation}
The numerical coefficients $c_s$ are constrained by
\begin{equation}
\sum_s c_s^2 =1 .
\end{equation}
One can optimize the coefficients $c_s$ to attain a better bound. 

Let us now argue why one cannot use rigorous methods to get a 
worst case condition number for domain wall fermions that is
better than the worst case condition number we obtained for
overlap fermions implemented by using $H$ (equation~(\ref{cond_nr_ineq})).

As we learned in the previous section, 
the key point is that one cannot rule out,
for arbitrary gauge field backgrounds, 
the existence of an eigenstate of $H_W$ with eigenvalue $h$ which
also is chiral and for which $|h|$ is very small. Suppose we have
such a state, call it $\psi_0$. This state is very special: since
$\gamma_5 H_W =D_W$, it is a simultaneous eigenstate of $H_W$ and
$D_W$. We choose:
\begin{equation}
\gamma_5 \psi_0 = \psi_0 ,\ \ \ H_W \psi_0 = D_W\psi_0 = h\psi_0 .
\end{equation}

Obviously, only a very special gauge background could accommodate
a state like this, where, in addition, we want $|h|$ to be very small.
Actually, we also want $h$ to be negative. If we construct a state
$\Psi$ out of $\psi_0$ with coefficients $c_s$ as above, we easily
see that the action of $D$ produces a state of the same structure
with only the coefficients $c_s$ changed to $c^\prime_s$.
The action on the coefficients can be immediately read off as:
\begin{equation}
\pmatrix {c^\prime_1\cr c^\prime_2 \cr
\vdots\cr  c^\prime_{n-1}\cr c^\prime_n\cr}=
[ 1+h - M ]
\pmatrix {c_1 \cr c_2\cr 
\vdots\cr  c_{n-1}\cr c_n\cr}={\tilde M}\pmatrix {c_n \cr c_{n-1}\cr 
\vdots\cr  c_2\cr c_1\cr} 
\end{equation}
Here,
\begin{equation}
{\tilde M} = (1+h-M)S = \pmatrix { 0 & 0 & \dots &  0 & -1  & 1+h \cr
                        0 & 0 & \dots & -1 & 1+h &  0   \cr
                   \vdots & \vdots & \dots &\vdots &\vdots &\vdots \cr
                        -1 & 1+h & \dots & 0 &0 & 0 \cr
                       1+h & 0 &  \dots & 0 &0 & \mu \cr}
\end{equation}
The advantage of introducing ${\tilde M}$ is that it is hermitian and
obeys
\begin{equation}
[ 1+h - M ][ 1+h - M ]^\dagger = {\tilde M}^2
\end{equation}
We need to find the lowest eigenvalue of ${\tilde M}$ for $|h| <<1$
and to leading order in $\mu$. We are assuming, as usual, that we are
close to the chiral limit where $\mu=0$:
\begin{equation}
n|h|=-nh >>1
\end{equation}
For $\mu=0$ we know that the lowest eigenvalue is practically 
zero~\cite{trunc} (it is responsible for $|\det {\tilde M}|=|1+h|^n$ being
exponentially small). The eigenstate associated with this eigenvalue 
has the following structure: $c_i = N(1+h)^{i-1}$
(because $(1+h)c_{i} \approx c_{i+1}$ for a very small 
eigenvalue) ~\cite{trunc}. 
The normalization $N$ is therefore, at $n=\infty$ ~\cite{plbfirst}
\begin{equation}
N= \sqrt{{1\over {\sum_{i=0}^\infty (1+h)^{2(i-1)}}}} =
\sqrt{-h(2+h)} .
\end{equation}
Now, first order perturbation theory gives for the smallest (in absolute
value) eigenvalue
of ${\tilde M}$:
\begin{equation}
\lambda_{\rm min} ({\tilde M} )\approx -h(2+h)\mu,\ \ \
\sqrt{\lambda_{\rm min} (D^\dagger D) } \le |(2+h)h|\mu .
\end{equation}

Thus, if a state of type $\psi_0$ exists, and under
the additional assumption that $|h|<<1$, we obtain an approximate
bound
\begin{equation}
\sqrt{{1\over{||(D^\dagger D)^{-1}||}}} \le {{2\mu}\over{||H_W^{-1}||}} .
\label{temp}
\end{equation}
This leads to essentially the same situation as 
in the overlap alternative:
At the end of section V we concluded that one could not hope to be
able to prove a better general bound on the condition number there
than that of equation~(\ref{cond_nr_ineq}). Combining 
equation (\ref{temp}) with the known upper bound
$||D^\dagger D||^{1\over 2} \le (10+m)$ ~\cite{bounds} 
(in four dimensions, with $-2<m\le 0$), produces a bound similar to
equation~(\ref{cond_nr_ineq}). One cannot expect to get a rigorous 
upper bound on the 
condition number of the Dirac operator employed for domain wall 
fermions that is superior to that obtained for the overlap 
implemented by $H$. 

\section{Relating parameters}

When comparing the overlap alternative to domain wall fermions one
needs some criteria to relate the parameters in both models. In a
real QCD simulation the criteria should be that the parameters be
chosen so that similar physics is being described. This is not as
unambiguous as it first sounds, but clearly beyond this paper. 
Here we suggest the following set of criteria: We need to prescribe
relations between the number of extra fields in each case, the parameters 
$m$ in the Wilson Dirac operator in each case and the mass parameters
$\mu$ in each case.

The easiest is to connect the
integers $n$ in the two cases. This we do by requiring the error
in realizing the sign function in both cases to be the same in the
realistic case that $H_W^2$ has some small eigenvalues $|h|$, 
of order $10^{-2}$
or less. The error for domain wall fermions is of order
\begin{equation}
e^{-{1\over 2} n_{\rm dmf} |h|} .
\end{equation}
The error in the alternative proposed in this paper is of order
\begin{equation}
e^{-2 n |h|} .
\end{equation}
But the number of Dirac fields in the domain wall case is
$n_{\rm dmf}$ while in the overlap alternative it is $n_{\rm ovp} = 2n+1$
So, for practical values we should take:
\begin{equation}
n_{\rm dmf}\approx 2n_{\rm ovp} .
\label{ncomp}
\end{equation}
This ensures similar violations of chirality at $\mu =0$ 
for the gauge backgrounds where these violations are most significant. 
This match of the number of extra fields works in favor of the
overlap alternative. 

To match the parameters $m$ in both cases we focus on the
problematic gauge backgrounds for which one would need very large
numbers of extra fields to reproduce the sign function
correctly (or else, use projection techniques). 
We recall that if the parameter $m$ is the
same in the domain wall and overlap contexts the logarithm of the
transfer matrix of domain wall fermions and the hermitian
Wilson Dirac operator of the overlap both 
acquire eigenvalues very close to zero in 
the same (bad) gauge configurations~\cite{npblong}. 

What is left is to match the parameters $\mu$. This is more
difficult to do in a gauge field independent way: Let us
adopt the criterion that we want the quark propagator masses
to be the same in both cases, for the same gauge background,
and infinite number of extra fields in both cases. The parameter
$m$ has already been chosen to be the same. Denoting by $u$
the fourth root of the plaquette variable ~\cite{paulpeter}, 
we find, in $d$ dimensions, 
the following ``mean-field improved'' estimates for the quark
masses in each case:

For domain wall fermions, assuming $0<\mu<<1$, we get:
\begin{equation}
m_{\rm phys}^{\rm dwf} =-
{{[m+d(1-u)][2+m+d(1-u)]}\over u} \ \mu .
\end{equation}

For the overlap fermions we get~\cite{zfac}:
\begin{equation}
m_{\rm phys}^{\rm overlap} = {{2\mu}\over{1-\mu}} {{|m+d(1-u)|}\over u } .
\end{equation}
Tree level perturbation theory is obtained by setting $u=1$. Above,
we assume that $-2<m<0$. 
Setting the two physical masses equal to each other we obtain for
small $\mu$
\begin{equation}
\mu_{\rm overlap} = \mu_{\rm dwf} \left [ 1+{{m+d(1-u)}\over 2} \right ] .
\end{equation}
In a QCD simulation one may take $u\approx .875$, $m\approx -1.8$, giving,
roughly, $\mu_{\rm overlap}\approx .35 \mu_{\rm dwf}$. This might
indicate an advantage to domain wall fermions, because the slow-down
on inversion of $H$ or $D$
as a result of a small $\mu$ might be roughly the same for equal
$\mu$ parameters. In our simulations of the Schwinger model 
$u$ is closer to unity, $u\approx .95$ 

\section{Projection technique}

In practice, most of the
numerical problems have to do with the presence of small eigenvalues 
to $H_W^2$ for typical gauge backgrounds. In the older applications
of the overlap one used a projection method to deal with this
difficulty. The projection method trivially extends to the overlap
alternative presented here.
Suppose we have several low states $\psi_a$,$H_W \psi_a = h_a \psi_a$ with
so small $|h_a|$ values that the needed $n$'s to handle these are
too large. Define the orthogonal set of projectors 
\begin{equation}
P_a = \psi_a \psi_a^\dagger , \ \ \ \sum_a P_a = {\cal P} .
\end{equation}
Redefine $H_W$ in the overlap alternative by:
\begin{equation}
H_W \to H_W^P = (1-{\cal P}) H_W (1-{\cal P}) +\sum_a {\rm sign}(h_a ) P_a .
\end{equation}
This does not change the effective $\psi$ action at $n=\infty$;
no new questions about locality appear as a result of this
replacement. The replacement 
dramatically reduces the $n$ needed to get close to the $\mu=0$ chiral
limit. The shifted states are now perfectly 
represented because for any $n$ we have $\varepsilon_n (\pm 1) =\pm 1$.
Of course, calculating the $P_a$'s and acting with ${\cal P}$ comes
with a cost; it is expected that the cost is bearable because the
number of states one needed to project out was of the order of $20$
in practical simulations carried out so far in the quenched
approximation using the rational approximation ~\cite{scri}. 

The introduction of projectors complicates the calculation of the force
in hybrid Monte Carlo, but the complication is manageable~\cite{dynovl}.
Under an infinitesimal change of the background gauge fields 
we have:
\begin{equation}
\delta P_a = \delta \psi_a\psi^\dagger_a
+ \psi_a \delta \psi^\dagger_a .
\end{equation}

We make a phase choice so that $P_a \delta \psi_a=0$
($\psi_a$ is normalized; the phase choice eliminates the
component of $\delta \psi_a$ in the direction of $\psi_a$
and we ignore possible degeneracies for simplicity) and then obtain
\begin{equation}
\delta \psi_a =
{{1-P_a}\over {h_a - H_W}}  \delta H_W  \psi_a .
\end{equation}
The variation of the projector now follows from that of the state,
and the phase choice we made no longer has any effect:
\begin{equation}
 \delta P_a =
{{1-P_a}\over {h_a - H_W}} \delta H_W  P_a 
+ P_a \delta H_W^\dagger {{1-P_a }\over {h_a - H_W}} . 
\end{equation}
The variation of the projectors enters the force
always when acting on a vector.
To obtain the resulting vector
we need to compute the action of an inverse on a vector. If
the overlap Dirac operator is implemented by the rational method
of ref~\cite{ratioprl,scri}, the action of ${1\over {c+ H_W }}$ on the same
vector for some constant $c$ has already been evaluated. Using the
shift trick of reference~\cite{shifttrick} one can evaluate this new inversion
at practically no additional cost in operations.  

Of course, if somebody comes up with a replacement of $H_W$ that
has a larger gap around zero, 
it just plugs in simply into the overlap alternative. 

Projection methods when generalized to domain wall fermions need to address
the more complicated form the transfer matrix $T_W$. But, if we are willing
to depart somewhat from the clean structure we have seen above in the
overlap case a natural suggestion is to replace $H_W$ by $H_W^P$
in the hermitian domain wall operator $H_{\rm dwf}$. We expect this
to help because this would shift the small energy modes of $H_W$
elsewhere in the spectrum. This is not an exact procedure in terms
of the transfer matrix $T_W$, but the unit eigenstates of $T_W$
(conjugated by $\gamma_5$) are also zero eigenstates of $H_W$.
Viewing the replacement as a perturbation, we see that the troublesome
modes are the ones going to be most significantly affected. The main
point in choosing an approximate projection is to avoid dealing 
with the nonlocal structure of the transfer matrix $T_W$. Numerical
experience from early overlap days teaches us that dealing with $T_W$
directly is possible~\cite{npblong}, 
but it is also rewarding to deal with the sparse
matrix $H_W$ instead. It would be interesting to check this suggestion
out in practice.

\section{Advantages of the overlap alternatives}

From the point of view of the projection technique the overlap
alternative is certainly cleaner. But, if the proposal for domain
wall fermions we just made
works, we may still feel that the domain wall approach and the
overlap approach ended up too close to a draw to make it worthwhile
to investigate the overlap alternatives.

Let us now turn to what we feel are more clear advantages of the overlap
alternatives. These advantages have to do with the simplified 
structure of the effective
$\psi$ action and with the fact that the light fermion is so well
identified. Unlike for domain wall fermions, where the ``wave function''
for the light quark penetrates a sizable amount into the extra dimension,
here the light fermion is fixed as $\psi$. There is no better context
to show how this would impact numerical QCD than to discuss chiral
symmetry at $\mu=0$: 

Let us introduce new $\psi^\prime$ fields so that the action now involves
the non-hermitian overlap Dirac operator $D_o$:
\begin{equation}
\psi^\prime = \psi, \ \ \ \bar \psi^\prime = \bar\psi\gamma_5 .
\end{equation}
Introduce yet another fermionic degree of freedom $\xi$, but this time an
auxiliary one: it has no kinetic energy. Integrating out all the
fermions but $\psi$ and $\xi$ leaves the following effective action:
\begin{equation}
S_{\rm eff} (\psi^\prime ,\xi) = -\bar\psi^\prime {{1+V_n }\over 2 } 
\psi^\prime + \bar\xi\xi . 
\end{equation}
The matrix $V_n$ is given by
\begin{equation}
V_n = \gamma_5 \varepsilon_n (H_W )
\end{equation}
and obeys:
\begin{equation}
V_n^\dagger V_n = \varepsilon_n^2 (H_W ) .
\end{equation}
At $n=\infty$, for an invertible $H_W$, we see that $V\equiv 
V_\infty$ is unitary and that the $\bar\psi^\prime, \psi^\prime$
action is given by the overlap Dirac operator, $D_o$
\begin{equation}
D_o = {{1+V}\over 2} .
\end{equation}
It is well known that $D_o^{-1} -1$ is chiral ~\cite{trunc}. 

Define the physical fermion field $\psi_{\rm ph}$ by
\begin{equation}
\psi_{\rm ph} = \psi +\xi .
\end{equation}
This definition was chosen so that we get, at $n=\infty$, the required
subtraction of unity (contributed by the $\xi$ propagator) 
from the $\psi$ propagator. 
\begin{equation}
\langle \psi_{\rm ph} \bar \psi_{\rm ph} \rangle = D_o^{-1} -1 =
{{1-V}\over {1+V}} .
\end{equation}
Therefore, the effective action for the physical field is
\begin{equation}
S_{\rm eff}({\psi_{\rm ph}}) = -\bar\psi_{\rm ph} 
{{1+V}\over {1-V}} \psi_{\rm ph} .
\end{equation}
This effective action is chirally symmetric: The transformation
\begin{equation}
\psi_{\rm ph} \to e^{i\alpha \gamma_5 } \psi_{\rm ph},\ \
\bar\psi_{\rm ph} \to  \bar\psi_{\rm ph} e^{i\alpha \gamma_5 }
\end{equation}
leaves $S_{\rm eff}$ invariant on account of
\begin{equation}
e^{i\alpha \gamma_5 }{{1+V}\over{1-V}} e^{i\alpha \gamma_5 }=
{{1+V}\over{1-V}} .
\end{equation}
A continuum fermion operator with desired chiral properties is
now simply transcribed to the lattice by replacing the continuum
fermionic fields by $\psi_{\rm ph}$. Violations of chirality
as a result of finite $n$ can be traced quite explicitly.  

Perturbative calculations are an integral part of any procedure
that connects numerical QCD to physics. In the overlap alternative
one can restrict ones attention only to the $\psi$ fields; there is no
need to deal with higher dimensional propagators. One has an explicit,
relatively simple action, and calculations, although still harder than
in the Wilson case, are tractable. Indeed, 
there has been progress 
on perturbative calculations with overlap fermions recently  
~\cite{vicari} (superficially, the treatment of chiral
symmetry there may seem somewhat more involved then the one
presented above, but it is essentially the same as here). As far as 
we know there are no finite $n$ calculations yet, but we expect them
to be relatively manageable. 

The parallel calculations in the domain wall case are
extremely cumbersome, see for example~\cite{deltam}. 
In particular, when one deals with questions related
to chiral symmetry one seems to have to deal with the fermions in 
the five dimensional ``bulk''~\cite{fur_shamir}. 
This makes the finite $n$ effects in
the domain wall context harder to estimate quantitatively. 
All in all, when working with domain wall fermions, one pays
quite a price for having the two chirality components of the light
fermion widely separated, and the entire set of extra fermions
actively involved in the communication between them, be it because
of finite $n$
effects or be it because of a topologically nontrivial gauge
background. 

Another factor to recall which is in favor of the overlap has to do with
the parameter $m$. In the overlap this parameter is theoretically
restricted to the range $-2 < m < 0$. When the effects of the noisy
gauge background are taken into account one gets effectively
a positive additive contribution to $m$ (in the mean field
approximation it is $d(1-u)$) which forces one towards the $-2$
end of the range. Still, one does not need to go beyond $-2$, so
one is safe even for trivial gauge backgrounds. In the domain
wall case one also is driven towards $-2$, but now there is some
theoretical worry: For $m$ in the range $-2 < m < -1$ the expression
for the transfer matrix $T_W$, while still a hermitian matrix, no longer
stays positive definite for all gauge fields~\cite{trunc}. There are gauge
backgrounds for which the matrix has negative eigenvalues. This 
raises some concerns about the true phase
the lattice model is in. Even only the proximity of a phase
different from continuum QCD is a source of potentially large,
undesirable, numerical effects. 
Theoretically one would like to stay
in the range $-1<m<0$ for domain wall fermions, but this cannot
be achieved in QCD at present typical simulation parameters. In
principle, one could get into this range, but the needed gauge
couplings would have to be impractically weak. Thus, the overlap
alternative seems safer for the coarse lattices currently 
employed in numerical QCD. 

Our specific overlap proposal is a very simple
implementation of one rational approximation to the sign function.
Clearly, there are many variations possible and there is room for
more improvement. In terms of flexibility, the overlap alternative
of this paper is superior to the domain wall approach; and more 
flexibility opens more possibilities to increase numerical efficiency.

Up to this point our main conclusion is that the overlap approach based
on the matrix $H$ is superior to the one based on domain
wall fermions. Of course, we cannot rule out some surprises,
so a numerical check in the context of QCD is necessary. 

\section{$H_W$ operation counts}

Going back to our main motivation, the reason to introduce $H$ in
the first place was to have a more efficient implementation of the
action of the rationally approximated overlap Dirac operator. The
original direct implementation used a two level nested conjugate gradient
(CG) procedure and
a mass-shift trick~\cite{ratioprl,scri,shifttrick}
which makes operation counts almost $n$-independent. In the inner
CG the relevant condition number is that of $H_W$ and in the outer
CG the relevant condition number is roughly ${1\over \mu}$. We see
that our worst case bound for the condition number relevant 
to the single step CG for $H$ is roughly the product of the
previous two condition numbers. Thus, it seems that the new procedure
is an order $n$ slower than the one it came to improve on. 
In addition, if one is willing to increase the operation count
by a factor of $2$ in the old procedure, one can eliminate all
the storage of the extra fields, providing a factor of $n$ saving
in memory~\cite{modphysicsapaper}; in a computation limited by memory
bandwidth rather than processor performance this version might
turn out to be the best. 

It is important to mention that these 
considerations ignore the possibility of preconditioning
the algorithm for inverting $H$; 
for example, the structure of $H$ readily admits
standard red-black preconditioning. Other preconditioning methods 
might exist, exploiting the rather smooth structure in $s$-space
as evident from the continuum $s$-limit with no $s$-derivatives.
Moreover, the bounds on the condition number are not really saturated
very often, and super-linear convergence effects in the CG procedure
may change the dependence of the number of required iterations
on the condition number away from the theoretical limit. Still, it
seems that for very large $n$ the older approach will eventually win.
However, both the older and newer approach can be improved by projectors,
and this will limit the size of $n$ one really needs. Also, the mass-shift
trick is incompatible with red-black preconditioning, so this might work
in favor of the new approach. 

To get some feeling for what one would see in practice we again turn
to two dimensional QED, with a simple plaquette Wilson action
at $\beta=4$ on an $8\times 8$ lattice. We performed the calculations
necessary to obtain $\langle \bar \psi_{\rm ph} \psi_{\rm ph}\rangle $ using
either method and counted the number of $H_W$ operations required to
reduce the norm of the residual to $10^{-8}$.
We used $n=20$ but no preconditioning in the $H$-algorithm. We did use
the mass-shift trick in the older algorithm. In neither method did we
include projectors. This comparison ended up in favor of the older method
by a factor of roughly 2.5. This factor could be beaten by red-black
preconditioning, but we have not tried this out. 

The above direct comparison between the two overlap 
methods is easy because they should
produce exactly the same results at the same $n$ and $H_W$, 
gauge configuration by gauge configuration.
Essentially, these are just two different algorithms to do the same thing. 
A comparison to domain wall fermions is more difficult, because the
differences at finite $n$ are more substantial and there is uncertainty
about how to match the parameters. As far as operation counts
go, one action of $D$ in the
domain wall case counts roughly as $ n_{\rm dwf}$ $H_W$ actions.

\begin{figure}
\epsfxsize = 0.8\textwidth
\caption{ A comparison of the number of operations of $H_W$ 
for the inversion of the fermionic operator in three cases:
domain wall fermions, the direct implementation of the rational approximated
sign function and 
the higher dimensional implementation of the overlap Dirac operator.
}
\centerline{{\setlength{\epsfxsize}{6in}\epsfbox[20 20 600 600]{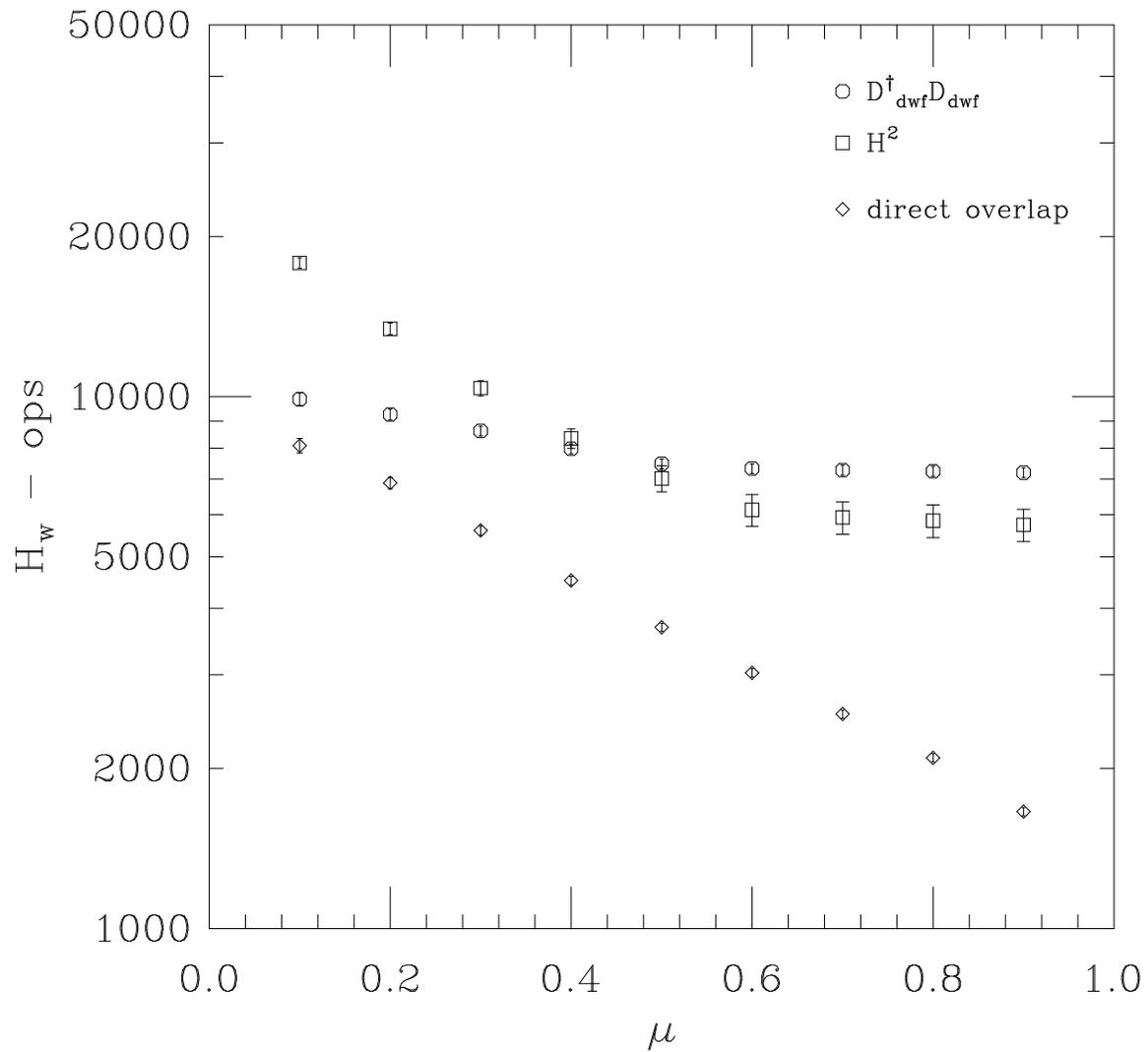}}}
\label{h_op}
\end{figure}

A plot of the average number of operations
of $H_W$ as a function of mass is shown for domain wall fermions, for the 
direct rational implementation and for the higher dimensional implementation
of the overlap Dirac operator in Fig.~\ref{h_op}. The data was obtained
from a sample of 20 configurations. The Wilson mass parameter was set
to $-1.5$ in all cases. Plotted on the vertical axis
is the number of operations needed for
a single inversion of the fermionic operator. For
the overlap Dirac operator simulations $n$ was set to $20$. The
plot also includes data for domain wall fermions, where 
$n$ was taken as $40$,  
in accordance with equation (\ref{ncomp}). 
We used the same parameter $\mu$ also for domain wall fermions because
the relation in eq. (85) holds only for $0< \mu << 1$ and becomes
totally inadequate at $\mu =1$. 

\section{Conclusions}

Let us summarize roughly the situation we were looking at when we began
this paper: Although an approach based on the overlap Dirac operator
looked theoretically cleaner, domain wall fermions were more attractive
numerically. Our analysis has led us to the conclusion that there
is no evidence that domain wall fermions have even a numerical advantage. 

In all cases we looked at, one faces a problem related to almost zero modes
of $H_W$. This requires large numbers of extra fields in order
to preserve chirality. It also
affects adversely the condition numbers. Whichever method we
use, the worst case condition numbers are a product of the inverses
of two main scale ratios: The first is the scale of the small eigenvalues
of $H_W^2$ divided by an upper bound of the order of 5-10 in lattice
units. The second scale ratio is the lattice physical quark mass squared 
divided by a number of order unity. Each small scale ratio slows down
inversion independently and the effect compounds in the worst case. 

Thus, as far as we can see, at the numerical level, there 
are no {\sl a priori} advantages to choosing domain wall fermions
over overlap fermions in the context of QCD. In both formulations
one faces similar numerical obstacles, and the overlap,
to say the least, does not fare any worse than domain wall fermions. 
At the analytical level we are convinced
that an approach based on the overlap (or any other efficient 
replacement of the overlap Dirac operator 
that might be found in the future) is superior at presently
attainable gauge couplings in numerical QCD. Perturbation theory
is more transparent to interpret and technically less complex in the
overlap version. The chirality violating effects associated with
the number of extra fields are much more explicit and therefore
their impact should be easier to trace through.

\acknowledgements

This research was 
supported in part by the DOE under grant \#
DE-FG05-96ER40559.


\begin{thebibliography}{99}
\bibitem{kaplan} D. B. Kaplan, Phys. Lett. {\bf B288}, 342 (1992).
\bibitem{slav} S. A. Frolov, A. A. Slavnov, Phys. Lett. {\bf B309}, 344
(1993).
\bibitem{plbfirst} R. Narayanan and H. Neuberger, Phys. Lett. {\bf B302},
 62 (1993).
\bibitem{domain} 
D. Boyanovsky, E. Dagotto, and E. Fradkin, Nucl. Phys. {\bf B285}, 340 (1987);
Y. Shamir, Nucl. Phys. {\bf B406}, 90 (1993).
\bibitem{ratioprl} H. Neuberger, Phys. Rev. Lett. {\bf 81}, 4060 (1998). 
\bibitem{ovldiracop}  H. Neuberger, Phys. Lett. {\bf B417}, 141 (1997);
Phys. Lett. {\bf B427}, 125 (1998).
\bibitem{trunc} H. Neuberger, Phys. Rev. {\bf D57}, 5417 (1998).
\bibitem{edwliu}  R.G. Edwards, U.M. Heller, R. Narayanan, hep-lat/9905028;
 K.F. Liu, S.J. Dong, F.X. Lee, J.B. Zhang, hep-lat/9909061.
\bibitem{columb} 
P. Vranas, Phys. Rev. {\bf D57}, 1415 (1998);
P. Chen et. al. Nucl. Phys. Proc. Suppl. {\bf 73}, 456 (1999);
T. Blum and A. Soni,  Phys. Rev. {\bf D56}, 174 (1997);
J.-F. Lagae, D. K. Sinclair, Nucl. Phys. Proc. Suppl. {\bf 73}, 450 (1999).
\bibitem{wilson} R.G. Edwards, U.M. Heller, R. Narayanan, 
Nucl. Phys. {\bf B535}, 403 (1998); Phys. Rev. {\bf D60}, 034502 (1999). 
\bibitem{scri} R. G. Edwards, U. M. Heller, and R. Narayanan,
Nucl. Phys. {\bf B540}, 457 (1998).
\bibitem{roburs} R. G. Edwards, U. M. Heller, hep-lat/0005002. 
\bibitem{fived} H. Neuberger, Phys. Rev. {\bf D60}, 065006 (1999).
\bibitem{bounds} H. Neuberger,  Phys. Rev. {\bf D61}, 085015 (2000).
\bibitem{npblong} R. Narayanan, H. Neuberger, Nucl. Phys. {\bf B443},
305 (1995).
\bibitem{paulpeter} G. P. Lepage and P. B. Mackenzie, 
Phys.Rev. {\bf D48}, 2250 (1993).
\bibitem{zfac} R. G. Edwards, U. M. Heller, and R. Narayanan,
Phys. Rev. {\bf D59}, 094510 (1999).
\bibitem{dynovl} A. Bode, U. M. Heller, R. G. Edwards, R. Narayanan,
hep-lat/9912043. 
\bibitem{shifttrick} A. Frommer, S. G\"usken, T. Lippert, B. N\"ockel,
and K. Schilling, Int. J. Mod. Phys. {\bf C6}, 627 (1995); 
B. Jergerlehner, hep-lat/9612014.
\bibitem{vicari} C. Alexandrou, E. Follana, H. Panagopoulos, E.
Vicari, hep-lat/0002010; C. Alexandrou, H. Panagopoulos, E.
Vicari, hep-lat/9909158;  M. Ishibashi, Y. Kikukawa, T.
Noguchi, A. Yamada, hep-lat/9911037. 
\bibitem{deltam} Y. Kikukawa, H. Neuberger, A. Yamada, 
Nucl. Phys. {\bf B526}, 572 (1998). 

\bibitem{fur_shamir} V. Furman and Y. Shamir, 
Nucl. Phys. {\bf B439}, 54 (1995).
\bibitem{modphysicsapaper} H. Neuberger, 
Int. J. Mod. Phys. {\bf C10}, 1051 (1999).
\end{thebibliography}
\end{document}